\documentclass[prl,twocolumn,showpacs,preprintnumbers,amsmath,amssymb]{revtex4}
\newcommand{\ket}[1]{\left | #1 \right\rangle}
\newcommand{\bra}[1]{\left \langle #1 \right |}

\newcommand{\qq}[1]{``#1"}
\usepackage{bm}
\usepackage{amsmath,amssymb,amsthm,times,graphics,graphicx,bm,xcolor}
\usepackage[normalem]{ulem}
\newtheorem{thm}{Principle}
\newtheorem{EN}{Theorem}
\newcommand{\A}{\textsc{a}}
\newcommand{\B}{\textsc{b}}
\newcommand{\M}{\textsc{m}}

\begin{document}

\title{Witnessing non-classicality beyond quantum theory}

\author{Chiara Marletto and Vlatko Vedral}
\affiliation{Clarendon Laboratory, University of Oxford, Parks Road, Oxford OX1 3PU, United Kingdom and\\Centre for Quantum Technologies, National University of Singapore, 3 Science Drive 2, Singapore 117543 and\\
Department of Physics, National University of Singapore, 2 Science Drive 3, Singapore 117542 and\\ISI Foundation, Turin}

\date{\today}

\begin{abstract}
We propose a general argument to show that if a physical system can mediate locally the generation of entanglement between two quantum systems, then it itself must be non-classical. Remarkably, we do not assume any classical or quantum formalism to describe the mediating physical system: our result follows from general information-theoretic principles, drawn from the recently proposed constructor theory of information. This argument provides the indispensable theoretical basis for recently proposed tests of non-classicality in gravity, based on witnessing gravitationally-induced entanglement in quantum probes.
\end{abstract}


\maketitle                           %display desi d

A class of experiments for detecting non-classicality in gravity has recently been proposed, \cite{SOUG, MAVE}. This has opened up an exciting possibility: quantum effects in gravity can be detected by probing indirectly the non-classicality of the gravitational interaction, through measuring the gravitationally-induced entanglement on {\sl two} quantum probes.
In this paper we focus on the theoretical foundations for experiments in this class. 

These experiments are based on the fact that if a system ${\bf M}$ (e.g. gravity) can entangle two quantum systems ${\bf Q_A}$ and ${\bf Q_B}$ (e.g. two masses) by local interactions, then ${\bf M}$ must be non-classical. 
By {\sl non-classical} we mean, informally, that the mediator ${\bf M}$ must have at least two variables that cannot be measured to arbitrarily high accuracy simultaneously (i.e. by the same measuring system). This is roughly what is meant by \qq{complementarity} in quantum theory, and it will be defined formally below.

If ${\bf M}$ obeys quantum theory, the above fact follows directly from theorems about Local Operations and Classical Communication (LOCC), \cite{HOR}: a decoherent channel cannot entangle two other quantum systems by local operations. In order to apply these theorems to the case of gravity, one has already to assume that it obeys quantum theory; an experiment based on this assumption would therefore test whether gravity has some coherence, so that some massive superpositions are allowed beyond certain scales. The arguments in \cite{SOUG} and related proposals \cite{TOM, TOM1} follow this line of argument and generalise it to apply to cases where the mediator's quantum observables are not measurable directly. However, the proposed experiments aim to probe cases (such as gravity) where the mediator ${\bf M}$ {\sl may or may not} obey quantum theory. Therefore, to provide an adequate theoretical foundation for the proposed tests, one needs to prove the above fact under less restrictive assumptions, without assuming quantum theory in full. A more general argument in this spirit was proposed in \cite{MAVE, MAVE1}, not assuming all the properties of quantum dynamics for the mediator. That argument, though, was still expressed via density operators, which are rooted in quantum theory. 

Here we provide a far more general argument, based on information-theoretic principles and the principle of locality {\sl only} (to be defined precisely later). We will also define generalisations of concepts such as {\sl non-classical} and {\sl observable} to describe the mediator, that are compatible with quantum theory's and general relativity's, but do not assume either of those theories. To this end, we resort to the principles of the constructor theory of information \cite{DEMA}, which provide a useful guide when neither quantum theory nor general relativity can be assumed. These principles allow one not to assume any specific dynamics for the mediator, therefore making our approach more general than the existing hybrid quantum-classical approaches, such as \cite{HAL, TER1}, where a generalised Hamiltonian dynamics is usually assumed. Our logic here is akin to that of Bell's theorem: just like Bell's theorem applies to a vast class of theories obeying general probabilistic assumptions, our theorem applies to a set of theories that obey general information-theoretic principles, also in the spirit of other principle-based arguments proposed to merge quantum theory and general relativity, such as \cite{FEY, DEW, MAVE2}.

\subsection{An example from quantum theory}

This section discusses an example where the mediator does obey quantum theory, to illustrate the logic of the general argument. We will use a qubit-based model, to elucidate the role of the mediator's non-classical degrees of freedom in the entanglement generation. In this example, the relevant degrees of freedom are just the $X$- and $Z$- components of a mediating qubit ${\bf Q_M}$; in the general argument, we will show that the mediator must have analogous properties to ${\bf Q_M}$, but we shall prove this  without assuming that ${\bf M}$ obeys quantum theory.  

Consider two qubits, ${\bf Q_A}$ and ${\bf Q_B}$; and a mediator qubit ${\bf Q_M}$. They all start their evolution in a product state; at a later time, ${\bf Q_A}$ and ${\bf Q_B}$ become entangled, via interacting each {\sl locally} with ${\bf Q_M}$. A simple model is an entangling gate acting between ${\bf Q_A}$ and ${\bf Q_M}$; then, a SWAP gate between ${\bf Q_M}$ and ${\bf Q_B}$. Assuming that $\ket{0}$ is the $+1$-eigenstate of the Z component of each qubit, an example of this entangling process is:

\begin{eqnarray}
\ket{0}_A\ket{0}_M\ket{0}_B& \underbrace{\rightarrow}_{{\rm Bell_{AM}}} &\ket{B_{+0}}_{AM}\ket{0}_B\\\nonumber &\underbrace{\rightarrow}_{{\rm SWAP_{MB}}}& \ket{B_{+0}}_{AB}\ket{0}_M
\end{eqnarray}

\noindent where $\ket{B_{+0}}$ is one of the Bell states, describing two maximally entangled qubits. In order for ${\bf Q_A}$ and ${\bf Q_B}$ to become entangled via ${\bf Q_M}$, the latter must itself be entangled with ${\bf Q_A}$, at least to the same degree as qubits ${\bf Q_A}$ and ${\bf Q_B}$ are at the end of the protocol. This requires the mediator ${\bf Q_M}$ to engage other variables in its dynamical evolution, such as the $X$- and $Y$- components, which do not commute with its $Z$ component (that it is initially sharp). Our proposed argument will establish the existence of a generalised version of these incompatible variables as the signature of non-classicality of the mediator ${\bf M}$, without assuming that the latter obeys quantum theory.

We can see more clearly how the incompatible variables are engaged in the entanglement generation by considering the Heisenberg picture. 

Let $q_{x\alpha}$ denote an operator representing the $X$-component of qubit $\alpha$ (likewise for the $Y$ and $Z$ components). These operators are defined on the $2^{3}$-dimensional Hilbert space of the 3 qubits. We have $q_{z\alpha}q_{x\alpha}= {\rm i} q_{y\alpha}$, $q_{z\alpha}^2={id}$ and likewise for all the other components, while components of different qubits commute. If the gate $U(t_n)$ operates between time $t_n$ and $t_{n+1}$, we shall denote by
\begin{equation}
O_{\alpha}(t_{n+1})=U(t_n)^{\dagger}O_{\alpha}(t_{n})U(t_n)
\end{equation}
the operator representing the observable $O$ of system ${\alpha}$ after its action. The initial conditions are fixed by choosing particular values for $q_{x\alpha}(t_{0})$, $q_{y\alpha}(t_{0})$, $q_{z\alpha}(t_{0})$, for all $\alpha$'s, and by the Heisenberg state $\rho_H$. We assume that the initial conditions are expressed as $q_{z\A}(t_0)=Z\otimes id^{\otimes 3}\equiv q_{z\A}$, where $Z$ is a Pauli matrix, and so on. We choose the Heisenberg state to be $\rho_H=\ket{{\bf 0}}\bra{{\bf 0}}$, the +1 eigenstate of the operator $\frac{1}{2}(id+Z)^{\otimes 3}$.

The state of each qubit $\alpha$ at time $t$ is completely specified by at least two components, e.g.\ $\{q_{x\alpha}(t), q_{z\alpha}(t)\}$. 
In this picture, we can describe the entangling operation mentioned above, as causing the following dynamical evolutions on each of the three qubits:

\begin{eqnarray}
{\bf Q_A}: &\{q_{x\A}, q_{z\A}\} &  \rightarrow \{q_{z\A} 
q_{x\M}, q_{x\A}\} \rightarrow \{q_{z\A}q_{x\M}, 
q_{x\A}\}\nonumber\\
{\bf Q_M}: &\{q_{x\M}, q_{z\M}\}& \rightarrow \{q_{x\M}, 
q_{z\M}q_{x\A}\}  \rightarrow \{q_{x\B}, q_{z\B}
\} \nonumber \\
{\bf Q_B}: &\{q_{x\B}, q_{z\B}\}& \rightarrow \{q_{x\B}, 
q_{z\B}\} \rightarrow \{q_{x\M}, q_{z\M}q_{x\A}
\}\nonumber \;.
\end{eqnarray}

Here, the first column represents the initial value of the qubit's descriptors; the second column represents their values at time $t_1$, when qubits $A$ and ${\bf M}$ are entangled; the third represents the final value, where qubits $A$ and $B$ are entangled: this can be verified by considering the expected value of an entanglement witness, evaluated at that time. 

One can see that ${\bf Q_M}$'s two incompatible observables (its $X$ and $Z$ component) are both engaged in mediating, by local interactions, the quantum correlations between ${\bf Q_A}$ and ${\bf Q_B}$. The logic outlined here is widespread in quantum information and underpins protocols such as teleportation and entanglement swapping; but it is very useful to bear it in mind in view of our aim: we will establish that a general mediator ${\bf M}$ has to be non-classical, just like ${\bf Q_M}$, {\sl without assuming that ${\bf M}$ obeys quantum theory}.  This will entail showing that in order to entangle two qubits by interacting with each one individually, ${\bf M}$ must have degrees of freedom that, analogously to the $X$ and $Z$ components of ${\bf Q_M}$ in the above example, are incompatible with each other. All of these notions will now be formally defined in this more general scenario where quantum theory may not fully be obeyed by the mediator.

\subsection{The interoperability principle for information}

\noindent Here we introduce a constructor-theoretic principle, the {\sl interoperability principle for information}, \cite{DEMA}, and we express the principle of locality, which are the foundation of the argument we intend to propose. To this end, we will summarise the concepts of constructor theory needed in order to express those principles. 

{\bf States, Attributes and Variables}. When dropping the assumption that a specific dynamics holds for ${\bf M}$, we can still maintain other notions, such as a generalised notion of {\sl state} - which provides the full description of a physical system. We will assume that ${\bf M}$ obeys a theory endowed with a set of allowed {\sl states} for physical systems and a partition of the whole universe into subsystems. We will be concerned with physical systems on which transformations can be performed, called {\sl substrates}. 

An {\sl attribute} {\bf n} of a substrate is the set of all states where the substrate has a given property. A {\sl variable} is a set of disjoint attributes of a substrate. (Note that variables and observables differ: the attributes in a variable may not be distinguishable, as explained below). 

A variable $V$ is {\sl sharp} on a given system, with value $v$, if the system is in a state belonging to the attribute ${\bf v}$ in that variable. 

For instance: a qubit is a substrate; the set of all $+1$-eigenstates of a given projector is an attribute; that projector is sharp with value $1$ whenever the qubit is in any one of those states. 

{\bf Possible/impossible tasks.} A {\sl task} specifies a general physical transformation of a substrate, in terms of ordered pairs of input/output attributes. For example, the NOT task on the attributes ${\bf 0 }$, ${\bf 1}$ is written as $\{ {\bf 0}\rightarrow{\bf 1}\;,\;\;{\bf 1}\rightarrow{\bf 0}\}$.

\noindent A task is {\sl impossible} if the laws of physics impose a limit to how accurately it can be performed. Unitary quantum theory, for instance, requires the task of cloning sets of non-orthogonal quantum states to be impossible \cite{NOCLO}. Otherwise, the task is {\sl possible}: there can be arbitrarily good approximations to a {\sl constructor} for it, which is defined as a substrate that, whenever presented with the substrates in any of the input attributes of the task, delivers them in one of the corresponding output attributes, and, crucially, {\sl retains the property of being capable of performing the task}. 

{\bf Locality as a constraint on states.} A cardinal principle of constructor theory is the {\bf principle of locality}, which can be expressed as a strict constraint on the states of substrates, as follows:

\begin{thm}
{\bf Locality}. The state of a substrate is a description of it that satisfies two properties: 1) Any attribute of a substrate, at any given time $t$, is a {\sl fixed function} of the substrate's {\sl state}; 2) Any state of a composite substrate ${\bf S_1}\oplus {\bf S_2}$ is an ordered pair of states $(s_1, s_2)$ of ${\bf S_1}$ and ${\bf S_2}$, with the property that if a task is performed on ${\bf S_1}$ only, then the state of the substrate ${\bf S_2}$ is not changed thereby. 
\end{thm}

The principle of locality in this form is satisfied by quantum theory, but the states do not correspond to the density operators. This is manifest by considering quantum theory's Heisenberg picture, \cite{HADE}. In the Heisenberg picture, the state of a quantum system is the vector of the generators of its algebra of observables (which are dynamical variables). For instance, 
in the case of a single qubit -- in the notation introduced earlier -- its state is the vector of time-dependent components $\hat q\doteq (q_x(t), 
q_y(t), q_z(t))$; the fixed function is $Tr(\bullet \rho_0)$, where the dot stands for any appropriate set of Hermitean 
operators in the span of $\hat q$, and $\rho_0$ is some (fixed) Heisenberg state. 

Now, considering a two-qubit system, the state of each qubit $\alpha$ at time $t$ is completely specified by at least two components, e.g.\ $\{q_{x\alpha}(t), q_{z\alpha}(t)\}$. The state of the joint system is likewise reconstructed given all of the observables in the set $\{q_{x\alpha}(t)$, $q_{z\alpha}(t)\}$, because
\begin{equation}
U(t_{n})q_{x\alpha}(t_n)q_{z\alpha}(t_n)U^{\dagger}(t_{n})=q_{x\alpha}(t_{n+1})q_{z\alpha}(t_{n+1})
\end{equation}
by unitarity. This is why quantum theory satisfies the principle of locality as expressed above, considering the q-valued descriptors of the Heisenberg picture as states. These descriptors are local in that sense because they contain all the information about a system's non-trivial history. 

Note also that the principle of locality in this form implies no-signalling: for if the state of ${\bf S_2}$ does not change when a transformation on ${\bf S_1}$ is implemented, the empirically accessible attributes of ${\bf S_2}$ cannot change either, since, by the principle of locality, they are fully specified by a fixed function of that state, \cite{HADE, PAUL}. 
\bigskip

{\bf Information media.} One can provide a general information-theoretic characterisation of the mediator ${\bf M}$ in our argument, by resorting to the concept of { information medium}, \cite{DEMA}. An {\sl information medium} is a substrate with a set of attributes $X$, called {\sl information variable}, with the property that the following tasks are possible: 

\begin{equation}
\bigcup_{{x}\in X}\left\{({\bf x},{\bf x_0})\rightarrow ({\bf x},{\bf x}) \right\}\;, \label{kk}
\end{equation}

\begin{equation}
\bigcup_{{x}\in X}\left\{{\bf x}\rightarrow \Pi({\bf x}) \right\} \label{dd}
\end{equation}

\noindent for all permutation $\Pi$ on the set of labels of the attributes in $X$ and some blank attribute ${\bf x_0}\in X$.

\noindent The former task corresponds to \qq{copying}, or cloning, the attributes of the first replica of the substrate onto the second, target, substrate; the latter, for a particular $\Pi$, corresponds to a logically reversible computation (which need not require it to be realised in a physically reversible way).  So, an information medium is a substrate that can be used for classical information processing (but could, in general, be used for more than just that). For example, a qubit is an information medium with respect to any set of two orthogonal quantum states.  

\noindent {\bf The interoperability of information}. Any two information media (e.g. a photon and an electron) must satisfy the {\bf principle of interoperability}, \cite{DEMA}, which expresses the intuitive property that classical information must be copiable from one information medium to any other, irrespective of their physical details. Specifically:
 
\begin{thm}
If ${\bf S_1}$ and ${\bf S_2}$ are information media, respectively with information variable $X_1$ and $X_2$, their composite system ${\bf S_1} \oplus {\bf S_2}$ is an information medium with information variable $X_1\times X_2$, where $\times$ denotes the Cartesian product of sets. 
\end{thm}

This principle implies that the task of copying information variables (as in eq. \eqref{kk}) from one information medium to the other is possible. It also requires the possibility of performing computations on ${\bf S_2}$ without simultaneously affecting ${\bf S_1}$, otherwise it would not be possible to perform independent permutations of variables of ${\bf S_1}$ or ${\bf S_2}$. This property is guaranteed by the principle of locality, as expressed earlier.  

\bigskip

\noindent We can now express information-theoretic concepts such as measuring and distinguishing, without resorting to formal properties such as orthogonality, linearity or unitarity. This is the other key feature of constructor theory that will allow our argument to be independent of particular dynamical models. The variable $X$ is {\sl distinguishable} if the task

\begin{equation}
\bigcup_{{x}\in X}\left\{{\bf x}\rightarrow {\bf q_x} \right\} \label{dd1}
\end{equation}

\noindent is possible, where the variable $\{{\bf q_x}\}$ is some information variable. If the variable $\{{\bf x_0}, {\bf x_1}\}$ is distinguishable, we say that the attribute ${\bf x_0}$ is distinguishable from ${\bf x_1}$.  This notion of distinguishability allows one to generalise the orthogonal complement of a vector space: for any attribute ${\bf n}$ define the attribute $\bar {\bf n}$ as the union of all attributes that are distinguishable from ${\bf n}$. 

%When given an information observable $X=\{x_1, x_2, .., x_N\}$, one can consider the attribute $a_{X}=\bigcup_{x\in X}x$. The attribute $\bar\bar{a_{X}}$ is the generalisation of the span of a set of orthogonal states $x_1, x_2, ...x_N$ in quantum theory \cite{DEUMA}. 

\noindent An {\sl observable} is an information variable whose attributes ${\bf x}$ have the property that $\bar{\bar{\bf x}}={\bf x}$; this notion generalises that of a quantum observable. An observable $X$ is said to be {\sl sharp} on a substrate, with value $x$, if the substrate is in a state that belongs to one of the attributes ${\bf x}\in X$. 

A special case of the distinguishing task is the perfect measurement task: 

\begin{equation}
\bigcup_{{x}\in X}\left\{({\bf x}, {\bf x_0})\rightarrow ({\bf x}, {\bf p_x}) \right\} \label{PD}
\end{equation} where the first substrate is the \qq{source} and the second substrate is the \qq{target}. From the interoperability principle, it follows that the above task must be possible for any information variable. 

In the constructor theory of information, one can also define a generalisation of quantum systems, called {\sl superinformation media}, \cite{DEMA}. A superinformation medium is an information medium with at least two information observables, $X$ and $Z$, such that {\sl their union is not an information variable}. We shall call these observables {\sl incompatible}, borrowing the terminology from quantum theory, because a measurer of one must perturb a substrates where the other observable is sharp, \cite{DEMA}. Qubits are special cases of superinformation media, \cite{DEMA}: one can think of $X$ and $Z$ as two non-commuting observables, whose attributes cannot all be copied by the same cloner, because of the no-cloning theorem, \cite{NOCLO}.

{\bf Non-classicality}. In our argument we will aim at establishing that ${\bf M}$ has a lesser property: non-classicality. By a substrate being {\bf non-classical} we shall mean an information medium ${\bf M}$, with maximal information observable $T$, that has a variable $V$, disjoint from $T$ and with the same cardinality as $T$, with these properties:

\begin{enumerate}
\item{} There exists a superinformation medium ${\bf S_1}$ and a distinguishable variable $E=\{{\bf e_j} \}$ of the joint substrate ${\bf S_1}\oplus {\bf M}$, whose attributes ${\bf e_j}=\{(s_j, v_j)\}$ are sets of ordered pairs of states, where  $v_j$ is a state belonging to some attribute in V and $s_j$ is a state of ${\bf S_1}$;
\item{} The union of $V$ with $T$ is {\sl not} a distinguishable variable; \\
\item{} The task of distinguishing the variable $E=\{{\bf e_j} \}$ is possible by measuring incompatible observables of a {\sl composite superinformation medium} including ${\bf S_1}$, but impossible by measuring observables of ${\bf S_1}$ only.\\
\end{enumerate}

This generalises the property of quantum complementarity to the case where ${\bf M}$ may not have the full information-processing power as a quantum system. For, contrary to superinformation media, in non-classical substrates the variable $V$ may or may not be an information {\sl observable} - it may not be permutable or copiable; yet, its existence requires ${\bf M}$ to enable non-classical tasks on other superinformation media, such as establishing entanglement.

\subsection{The argument} 

We can now formulate our central argument using these information-theoretic tools and principles, under the following assumptions:

\begin{itemize}
\item{} The mediator ${\bf M}$ is an information medium with a maximal information observable $T$.
\item{} The two systems to be entangled, ${\bf Q_A}$ and ${\bf Q_B}$, are qubits. 
\end{itemize}

${\bf Q_A}$ and ${\bf Q_B}$ qualify as superinformation media, having at least two {\sl disjoint} maximal information observables, say their X and Z components, whose union is not an information observable. For simplicity, we will assume that all the information observables are binary: $T=\{{\bf t_0}, {\bf t_1}\}$; for the qubits, we have: $Z=\{{\bf z_1}, {\bf z_2}\}$ and $X=\{{\bf x_+}, {\bf x_-}\}$, where $X$ and $Z$ represent the $X-$ and $Z-$ component of each qubit, respectively.  

We now proceed to demonstrate our main result: 
\begin{EN}
If ${\bf M}$ can entangle ${\bf Q_A}$ and ${\bf Q_B}$, by locally interacting with each, then ${\bf M}$ is non-classical. 
\end{EN}

To prove this result, we will follow this logic. First, the interoperability principle implies that the following task is possible: to copy any of the observables of $Q_{\alpha}$ onto the observable $T$ of the mediator ${\bf M}$, via some interaction. We will assume that, by coupling $M$ locally with each of the qubits via that same interaction, it is possible to prepare them in one of two orthogonal maximally entangled states. By locality, this must be implemented by repeating two elementary steps: first performing a task on ${\bf Q_A}\oplus{\bf  M}$ and then on ${\bf M}\oplus {\bf Q_B}$. We will run the argument assuming entanglement is obtained via these two elementary steps, as it is straightforward to generalise to the case where a repetition of the two steps is required. Upon performing the former task, ${\bf M}$ is prepared in one of two attributes, by the principle of locality. These attributes, we shall argue, must belong to a binary variable $V$ satisfying the non-classicality conditions, just like the descriptors of the qubit $Q_M$ in our qubit-based example.

\noindent We proceed now with presenting the argument in full. We first establish the fact that ${\bf Q_A}\oplus{\bf M}$ must have an additional variable $E$ (generalising a set of entangled states), as in the first condition for non-classicality. 

\begin{itemize}
\item{} Given the Principle of Interoperability, the task of measuring the observable $X$ of one of the qubits, using the mediator ${\bf M}$ as the target, must be possible: 

\begin{eqnarray}
T_{M}\doteq \{({\bf z_0},{\bf t_0})\rightarrow({\bf z_0},{\bf t_0}),\\ \nonumber
({\bf z_1},{\bf t_0})\rightarrow ({\bf z_1},{\bf t_1}) \}\;,\label{EQU0}
\end{eqnarray}

where the first slot represents one of the qubits; the second slot represents the mediator.  In the limit of weak field, relevant for the tests in \cite{SOUG, MAVE}, one can think of ${\bf z_0}$ and ${\bf z_1}$ as two distinct locations of a mass; and of ${\bf t_0}$ and ${\bf t_1}$ as two distinguishable configurations of the gravitational field, induced by two different mass distributions ${\bf z_0}$ and ${\bf z_1}$. It is also possible to think of ${\bf t_0}$ and ${\bf t_1}$ as two distinguishable spacetime geometries, solutions of Einstein's equations for the two different mass distributions, as prescribed by general relativity, \cite{CHRO}.

\item{} If the experiment is successful in entangling ${\bf Q_A}$ and ${\bf Q_B}$, the following task must also possible:

\begin{eqnarray}
T_{E}\doteq \{({\bf x_+},{\bf t_0},{\bf x_+})\rightarrow{\bf e_{++}}, \nonumber \\
({\bf x_-},{\bf t_0},{\bf x_+})\rightarrow{\bf e_{-+}}\}\;\label{EQ1}
\end{eqnarray}

where $B\doteq\{{\bf e_{++}}, {\bf e_{-+}}\}$ is an information variable of ${\bf Q_A}\oplus {\bf M}\oplus {\bf Q_B}$ whose attributes correspond to two orthogonal, maximally entangled states of the two qubits. These attributes can be distinguished by measuring the observables of ${\bf Q_A}$ and ${\bf Q_B}$ only: specifically, let us assume that ${\bf e_{++}}$ is a maximally entangled state where both $X_A$, $X_B$ and $Z_A$, $Z_B$ are maximally correlated; while in ${\bf e_{-+}}$ the observables $X_B$ and $X_B$ are maximally correlated, while $Z_A$ and $Z_B$ maximally anti-correlated. The proposed experiments \cite{SOUG, MAVE} would show that the  task $T_E$ is possible, upon successfully generating entanglement between the probes ${\bf Q_A}$ and ${\bf Q_B}$. 

\item{} Assume also that the constructor for the task $T_E$ is the same as the constructor for the task $T_M$, so these two tasks can be performed jointly by the same interaction. In the case of the experiment with gravity, the constructor is the gravitational interaction between a mass and the gravitational field, initially prepared in some classical configuration, $t_0$. Also, we assume that $T_E$ is performed without ${\bf Q_A}$ and ${\bf Q_B}$ interacting directly. By the Principle of Locality, it must be performed in at least two steps; the first only involving ${\bf Q_A}$ and ${\bf M}$, the second only ${\bf M}$ and ${\bf Q_B}$. 

In the first step, this task is performed on ${\bf Q_A}\oplus {\bf M}$: 

\begin{eqnarray}
T_1\doteq\{({\bf x_+},{\bf t_0},{\bf x_+})\rightarrow ({\bf s_{+0}},{\bf x_+}), \\\nonumber
({\bf x_-},{\bf t_0},{\bf x_+})\rightarrow ({\bf s_{-0}},{\bf x_+})\}\;;\label{E1}
\end{eqnarray}

In the second step, this other task on ${\bf M}\oplus {\bf Q_B}$ is performed:

\begin{eqnarray}
T_2\doteq\{({\bf s_{+0}},{\bf x_+})\rightarrow {\bf e_{++}}, \\ \nonumber
({\bf s_{-0}},{\bf x_+})\rightarrow {\bf e_{-+}}\}\;.\label{E2}
\end{eqnarray}

From the possibility of task $T_1$, we see that the substrate ${\bf Q_A}\oplus {\bf M}$ has a variable: $E= \{{\bf s_{+0}}, {\bf s_{-0}}\}$. We now proceed to establish its properties, to show that ${\bf M}$ is non-classical. 

\item{}First, note that $E$ is a {\sl distinguishable variable}, because it can be mapped one-to-one onto two distinguishable attributes of the qubits, ${\bf e_{\alpha\beta}}$, via task the $T_2$. 

\item{} By the Principle of Locality, there are states $\hat q_A^{\alpha 0}$ of ${\bf Q_A}$ and $m^{\alpha 0}$ of ${\bf M}$ such that each of the attributes in the variable $E=\{{\bf s_{\alpha0}}\}$ is a fixed function of $(\hat q_A^{\alpha 0}, m^{\alpha 0})$ (where $\alpha$ takes values in $\{ +,-\}$). Here $\hat q_A^{\alpha 0}$ is a vector of q-numbers representing the three components of the qubit; while $m^{\alpha 0}$ is some state describing ${\bf M}$, whose properties we wish to establish.
\end{itemize}

\noindent We proceed now to establish the properties of the set of attributes $V\doteq\{\{m^{\alpha 0}\}\}$, to show that ${\bf M}$ is non-classical. 

\begin{enumerate}
\item{{\bf Condition 1 for non-classicality.}} First we prove that the set $V=\{ \{m^{+0}\}, \{m^{-0}\}\}$ is a binary variable (i.e., a set of two disjoint attributes).

{\sl Proof. } The Principle of Locality requires the states ${\bf e_{++}}$ to be a fixed function of the states describing ${\bf Q_B}$ and ${\bf M}$ prior to performing $T_2$; likewise for ${\bf e_{-+}}$. Specifically, let us denote by $\hat q_B^{++}$ the state of ${\bf Q_B}$ {\sl after} performing $T_2$, when the overall attribute is ${\bf e_{++}}$; and by $\hat q_B^{-+}$ the state of ${\bf Q_B}$ {\sl after} performing $T_2$, when the overall attribute is ${\bf e_{-+}}$. By the Principle of Locality, $\hat q_B^{\alpha +} = H(\hat q_B, m^{\alpha 0})$, where $H$ is some (well-behaved) function and $\hat q_B$ is a (q-numbered) state describing ${\bf Q_B}$ when it is in its initial attribute ${\bf x_+}$ (where $X$ is sharp with value $x_+$). 

We now use this fact to argue that $m^{+ 0}\neq m^{- 0}$. First, $e_{++}$ is distinguishable from $e_{-+}$ only by measuring observables of {\sl both} ${\bf Q_A}$ and ${\bf Q_B}$. Also, prior to performing $T_2$, the attributes  $({\bf s_{+0}},{\bf x_+})$ and $({\bf s_{-0}},{\bf x_+})$, though overall distinguishable, are not distinguishable by measuring observables of ${\bf Q_B}$ jointly with observables of ${\bf Q_A}$. This is because ${\bf Q_B}$ is still in the same initial state $\hat q_B$ where the observable $X$ is sharp with value $x_+$. 

Thus, the state $m^{+0}$ must be different from $m^{-0}$, as the dependence on $m^{\alpha 0}$ makes each of the $\{\hat q_B^{\alpha +}\}$ different from $\hat q_B$. Hence the set of attributes $V=\{ \{m^{+0}\}, \{m^{-0}\}\}$ is a variable (a set of disjoint attributes), with the same cardinality as $T$. Thus, ${\bf M}$ satisfies condition 1 for non-classicality. 

\item{{\bf  Condition 2 for non-classicality.}} Next, we prove that the attributes in $V$ are not distinguishable from, and do not overlap with, those in $T$. 

{\sl Proof.} Given that the task $T_2\cup T_M$ is possible (i.e., the two tasks are performed by the same constructor), each attribute $\{m^{\alpha 0}\}$ is not distinguishable from either ${\bf t_0}$ or ${\bf t_1}$. If it were, the attributes ${\bf x_+}$ and ${\bf x_-}$ of the qubit ${\bf Q_A}$ would be distinguishable from some of the ${\bf z}$'s, contrary to the assumption that ${\bf Q_A}$ is a superinformation medium.  For the same reason, $m^{\alpha 0}\not \in {\bf t_0}$ and $m^{\alpha 0}\not \in {\bf t_1}$. Therefore, ${\bf M}$ satisfies condition 2) for non-classicality. 

\item{{\bf Condition 3 for non-classicality.}} We note that the variable $V$ cannot be distinguished by measuring observables on ${\bf Q_A}$ only; it can be distinguished only by jointly measuring the complementary observables $X_A$ and $Z_A$ and $X_B$ and $Z_B$ on the superinformation medium ${\bf Q_A}\oplus {\bf Q_B}$. Hence, ${\bf M}$ satisfies also condition 3) for non-classicality.
\end{enumerate}
This concludes our proof that ${\bf M}$ is non-classical. 

\subsection{Discussion} 

What could the attributes $\{m^{\alpha+}\}$ in the variable $V$ be? Could they, for example, correspond to two different statistical mixtures of ${\bf M}$'s classical observable $T$, ${\bf t_0}$ and ${\bf t_1}$? The answer is no. This is because by performing the task $T_2$ and subsequently measuring observables of ${\bf Q_A}$ and ${\bf Q_B}$ jointly, one reveals entanglement between ${\bf Q_A}$ and ${\bf Q_B}$, which did not exist before the interaction between ${\bf Q_B}$ and ${\bf M}$. The correlations between observables of ${\bf Q_B}$ and those of ${\bf Q_A}$ after performing $T_2$ must be contained in the state $m^{\alpha +}$'s, given the locality principle: they are absent in ${\bf Q_B}$ before the interaction with ${\bf M}$, via $T_2$, while they are present in ${\bf Q_B}$ after performing $T_2$, when its state becomes dependent on $m^{\alpha +}$. Informally, the variable $V=\{\{m^{\alpha+}\}\}$ has at least the same information-carrying capacity as the q-number-valued states of the qubit ${\bf Q_B}$, because it contains all the correlations that are proper of an entangled qubit, as later confirmed by measurements of ${\bf Q_B}$. By Bell's theorem, $m^{\alpha +}$ cannot be a statistical mixture of ${\bf t_0}$ and ${\bf t_1}$, because, if it were, it would provide a local hidden variable model for quantum entangled states of ${\bf Q_A}\oplus {\bf Q_B}$. This argument, therefore, shows that collapse models, which would predict ${\bf M}$ to be in a statistical mixture of the observable $T$, are incompatible with observing entanglement. 

Thus the $\{m^{\alpha+}\}$ are not hidden variables, or \qq{beables}. They generalise the q-valued descriptors of what can dynamically change in a quantum system -- the descriptors of the quantum Heisenberg picture. In this sense, they are closer to the observables as conceived by von Neumann in his argument to rule out hidden variable models, \cite{VON}. Indeed, our argument could be understood as a first step towards generalising Bell's theorem to inferring non-classicality of systems, like ${\bf M}$, that can be used to {\sl assist locally} the violation of Bell inequalities on two other quantum systems, but need not have a full set of observables like a quantum system and therefore cannot violate Bell Inequalities directly.  

Another interesting point is that the variable $V$ may or may not be an information variable. If ${\bf M}$ were a qubit entangled with ${\bf Q_A}$, $V$ could not be an information variable, (otherwise we would be able to locally distinguish one entangled state from another just by measuring that information variable on ${\bf M}$). But given that system ${\bf M}$ may not obey quantum theory, so we must leave this possibility open. Note also that ${\bf M}$, although capable of working as a faithful channel for creating entanglement between the two qubits, may not have the full repertoire of operations such as preparation and measurement as a superinformation medium, let alone as a qubit. 

Our argument does not commit to any particular formalism to describe ${\bf M}$ and its interaction with the two qubits, in contrast with the thorough analysis of the gravity experiment presented in \cite{MAVE, SOUG, MAVE1, SOUG2}, where specific dynamical models are assumed. But how general are the principles we assumed? 
The interoperability principle holds in any physical theory that allows for measurements and observables - whose existence is a prerequisite for any physical theory to be testable. Therefore, it is a robust principle.  The principle of locality in the form discussed here is also satisfied by both quantum theory and general relativity. In \cite{PAUL} it also is proven that all theories based on $1:1$, no-signalling dynamics satisfy this principle of locality, thus making it a remarkably general property. This more general argument is of the essence for the witness of non-classicality to hold irrespective of whether the mediator is assumed or not to obey specific quantum models. It is the essential theoretical underpinning for experiments assessing the quantisation of gravity in full generality, where one cannot assume that gravity obeys a specific quantum model prior to the experiment. It ensures that if entanglement is observed, then {\sl all} classical models for gravity, obeying our general principles, are ruled out. This is similar to Bell's theorem, which ensures that if Bell's inequalities are violated by a given theory, then all local hidden variable models for that theory are ruled out. 
Our argument could have interesting implications for quantum gravity theories: it would be interesting to understand what the mediator and its non-classical variable $V$ are in each of the quantum gravity models that have been proposed, particularly non-perturbative ones. One could also consider lifting the assumption that ${\bf Q_A}$ and ${\bf Q_B}$ are qubits, and proceed with the general theory of superinformation media, \cite{HADE, DEMA, MA}, where entanglement is treated as {\sl locally inaccessible information}. We conjecture that even in this case, the degree of locally inaccessible information on ${\bf Q_A}$ and ${\bf Q_B}$ can be expressed formally as a function of the degree of non-classicality of ${\bf M}$, generalising the formal relation existing in quantum theory between non-classicality of the mediator and degree of entanglement, \cite{BOH}. 

This argument is effective to derive predictions in areas where specific dynamics cannot be assumed, going beyond current approximation schemes (such as open-system dynamics) or hybrid dynamical approaches (see e.g. \cite{TER1}).  The information-theoretic principles of constructor theory we used here provide a fruitful alternative to dynamics and initial conditions, useful to construct a bridge towards new theories of physics. 
In this paper we have demonstrated the first experimental application of this powerful approach.

\textit{Acknowledgments}: The authors thank David Deutsch and Anicet Tibau Vidal for providing sharp criticism on this manuscript. CM thanks the Eutopia Foundation, the Templeton World Charity Foundation, the John Templeton Foundation, and the FQXi. VV thanks the National Research Foundation, Prime Minister's Office, Singapore, under its Competitive Research Programme (CRP Award No. NRF- CRP14-2014-02) and administered by Centre for Quantum Technologies, National University of Singapore. This research was also supported by grant number (FQXi-RFP-1812) from the Foundational Questions Institute and Fetzer Franklin Fund, a donor-advised fund of Silicon Valley Community Foundation. This publication was made possible through the support of the ID 61466 grant from the John Templeton Foundation, as part of the The Quantum Information Structure of Spacetime (QISS) Project (qiss.fr). The opinions expressed in this publication are those of the authors and do not necessarily reflect the views of the John Templeton Foundation.

\end{document}